\documentclass[lettersize,journal]{IEEEtran}
\usepackage{amsmath,amsfonts}
\usepackage{algorithmic}
\usepackage{algorithm}
\usepackage{array}
\usepackage[caption=false,font=normalsize,labelfont=sf,textfont=sf]{subfig}
\usepackage{textcomp}
\usepackage{stfloats}
\usepackage{url}
\usepackage{verbatim}
\usepackage{graphicx}
\usepackage{cite}
\usepackage{siunitx}
\usepackage{multirow}
\usepackage{booktabs}
\usepackage{xcolor}
\usepackage{tikz}

\usepackage{pgfplots}
\pgfplotsset{compat=1.18}

\usetikzlibrary{arrows.meta,positioning}
\usepackage[colorlinks=true,linkcolor=blue,urlcolor=blue,citecolor=blue]{hyperref}

\newcommand{\nrx}{N^{\mathrm{Rx}}}

\newcommand{\Sph}{\mathbb{S}^{2}}
\newcommand{\R}{\mathbb{R}}
\newcommand{\Hnull}{\mathcal{H}_{0}}
\newcommand{\Halt}{\mathcal{H}_{1}}
\newcommand{\Dobs}{\mathcal{D}}

\newcommand{\indicator}{\mathbb{I}}

\definecolor{darkgreen}{rgb}{0.0, 0.5, 0.0}
\definecolor{darkorange}{rgb}{0.5, 0.3, 0.1}
\definecolor{orange}{rgb}{1,0.5,0}

\newcommand{\yas}[1]{\textcolor{black}{#1}}

\title{Receiver-Surface Hit Patterns via Legendre Approximation for Molecular Signal Detection}

\author{Yasin Bastug$^*$, Erencem Ozbey$^*$, and H. Birkan Yilmaz
    \thanks{Y. Bastug, E. Ozbey, and H. B. Yilmaz are with NETLAB, Department of Computer Engineering, Bogazici University, Istanbul, 34342, Turkiye (e-mail: birkan.yilmaz@bogazici.edu.tr).}
    \thanks{$^*$ These authors contributed equally to this work.}
}

\usepackage[normalem]{ulem}
\begin{document}

\maketitle

\begin{abstract}
Detecting whether a transmitter is actively communicating with a receiver is a fundamental problem in molecular communications. A spherical receiver may measure not only the number and arrival times of absorbed molecules, but also their absorption locations on the receiver surface. These locations contain a directional signature that is lost in count-only detection. In this letter, we develop a molecular signal detector based on a Legendre polynomial expansion of the receiver-surface hit density and extend it to a Legendre-based Viterbi sequence detector. By exploiting the surface-level directional signature of molecular arrivals, the proposed detectors provide geometry-aware, efficient and better-performing alternatives to count-only detection.
\end{abstract}


\section{Introduction}
\label{sec_introduction}

Molecular Communication via Diffusion (MCvD) is a communication paradigm in which transmitter (Tx) nodes encode information by releasing molecules into a fluidic environment, while receiver (Rx) nodes observe the molecules after their random propagation through the medium \cite{nanoscale_brownian}. Since molecule transport is governed by Brownian motion, the received signal depends probabilistically on the communication geometry, diffusion coefficient, number of released molecules, and relative Tx-Rx position~\cite{Kadloor2012_BrownMotion,kuran2020survey}. 

MCvD appears naturally in biological systems and has been explored for applications such as targeted drug delivery, in-body health monitoring, tumor detection, and cooperative nano-networking~\cite{nakano2012mcvd}, where the receiver often must first decide whether a meaningful molecular signal is present before decoding or localizing it. Thus, the Rx must be able to distinguish transmitter-induced molecule arrivals from background molecules, environmental noise, or molecules released by unrelated sources \cite{jing2023isi, optimal_receiver}.

A common receiver model in MCvD is the perfectly absorbing spherical receiver \cite{3d_channel_characteristics_via_absorbing_receiver}. Once a molecule hits the Rx surface, it is absorbed and removed from the environment. The total number of absorbed molecules over an observation interval contains distance-dependent information through the hitting probability function. However, a count-only observation discards a major part of the received signal. The absorption locations on the spherical surface form a directional pattern: molecules emitted by a Tx are not uniformly absorbed on the receiver surface, but concentrate according to the Tx direction, Tx-Rx distance, diffusion dynamics, and Rx geometry.

As a detection statistic, this \yas{letter} focuses on the receiver-surface hit pattern. Instead of treating the absorbed molecules only as a count, we model the angular distribution of the molecule absorption points. Due to the nature of Brownian motion, molecules absorbed shortly after a Tx emission form a directional surface pattern, whereas later ISI arrivals become more diffused, as illustrated in Fig.~1. Consequently, a communicating Tx generates a structured, direction-dependent surface distribution, while background molecules or unrelated molecular activity may not exhibit the same spatial structure. \yas{Detection therefore compares the angular-time distribution of noise alone against that of noise plus an active signal.}

Exact characterization of the receiver-surface distribution usually requires solving diffusion equations with absorbing boundary conditions. Although analytical expressions may exist in certain scenarios \cite{hitting_exterior, joint_distribution}, they are often difficult to use directly in practical likelihood-based detection. To overcome this issue, we use finite Legendre polynomial expansions that approximate the axially symmetric angular hit density induced by a Tx. Since the relevant angular variable is the cosine similarity between the Tx direction and the molecule absorption direction, Legendre polynomials provide a natural basis for representing the spherical hit pattern.

In addition to the memoryless Legendre likelihood-ratio detector, we also design a Legendre-Viterbi detector, where the Legendre-based angular-time likelihood is used within sequence detection. This version combines receiver-surface spatial information with Viterbi memory, enabling a direct comparison with the count-based Viterbi detector \cite{receiver_design, viterbi1967error, maximum_likelihood}.

The contributions of this \yas{letter} are summarized as follows:

\begin{itemize}
    \item We develop a Legendre polynomial representation of the angular hit distribution, capturing the directional and time-dependent structure of absorption patterns.

    \item We derive three Legendre-based detectors that combine molecule count, surface location, and arrival time information in memoryless and Viterbi-based schemes.

    \item We show that incorporating receiver-surface angular information improves detection over count only approaches, particularly in low-observation regimes, while offering different performance-complexity tradeoffs.
\end{itemize}

%%%%%%%%%%%%%%%%%%%%%%%%

\begin{figure}[t]
    \centering
    \includegraphics[width=\linewidth]{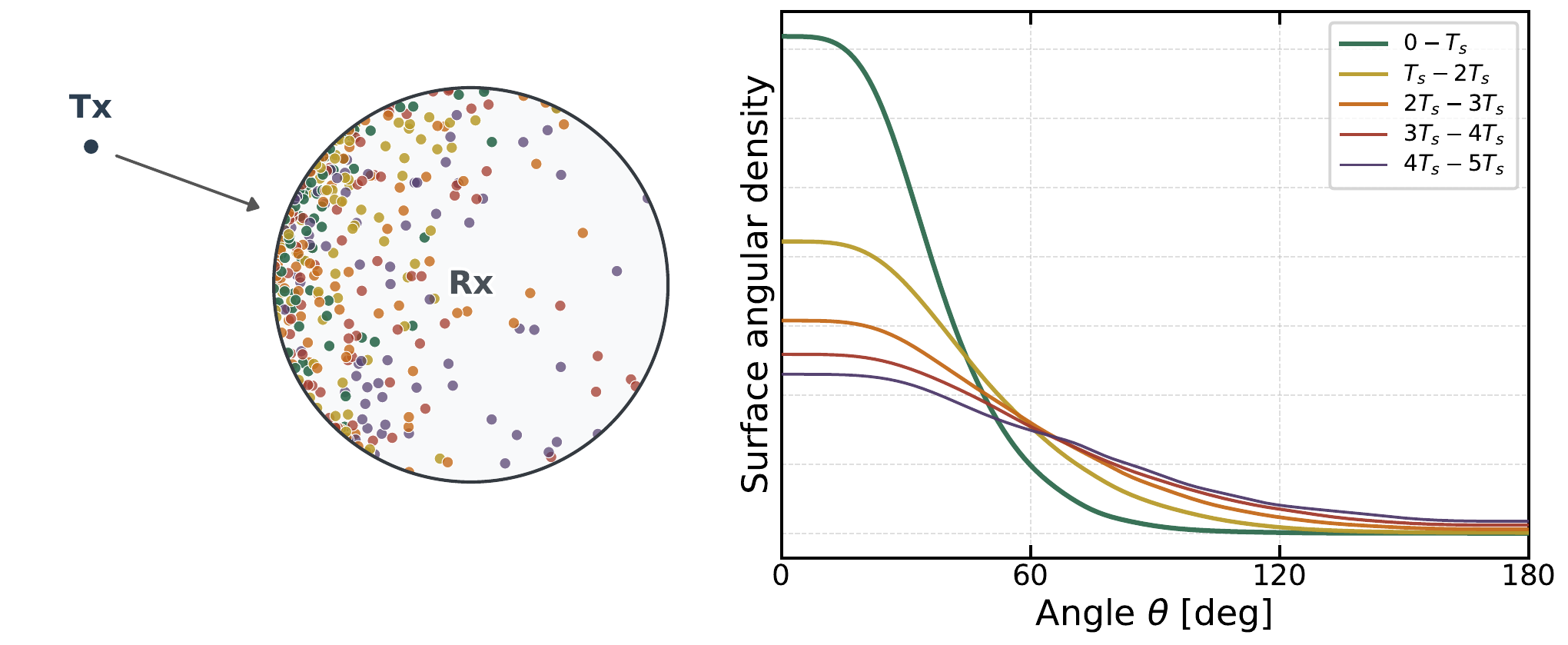}
    \caption{Surface hit pattern from a single Tx emission over five consecutive symbol intervals. Early arrivals are concentrated near the Tx-facing side of the Rx, while later ISI arrivals spread over the surface and become more uniform. The right panel shows the angular density versus \(\theta\), the angle between the Tx direction and the absorption point, represented through \(z=\cos\theta\).} 
\label{fig:detection_scenarios}
\end{figure}

\section{System Model}
\label{sec_system_model}

We consider a molecular communication system employing Binary Concentration Shift Keying (BCSK) modulation. 
Our core setup consists of a single point Tx and a single perfectly absorbing spherical Rx within an unbounded 3-dimensional environment. The Rx is centered at the origin and has radius \(r\). The Tx position is denoted by \(\mathbf{s}\in\R^3\), and is assumed to be known or localized by the Rx. Its distance from the Rx center and its direction are defined as
\begin{equation}
    d=\|\mathbf{s}\|, 
    \qquad
    \mathbf{a}=\frac{\mathbf{s}}{\|\mathbf{s}\|},
    \qquad
    \mathbf{a}\in\Sph .
    \label{eq:tx_position}
\end{equation}
Here, \(d\) is the Tx-Rx center distance, \(\mathbf{a}\) is the unit direction from the Rx center toward the Tx and $\mathbb{S}^2 = \{\mathbf{x}\in\mathbb{R}^3 : \|\mathbf{x}\|_2 = 1\}$. Molecules propagate in an unbounded fluid medium according to Brownian motion with diffusion coefficient \(D\). Once a molecule hits the Rx surface, it is absorbed and removed from the environment. In such scenarios, the probability that a molecule being absorbed by time \(t\) is
\begin{equation}
    F_{\mathrm{hit}}(t,d,r)
    =
    \frac{r}{d}\operatorname{erfc}\!\left(
    \frac{d-r}{\sqrt{4Dt}}
    \right).
    \label{eq:fhit}
\end{equation}

For a signal duration $T_s$ the capture probability is $F_{\mathrm{hit}}(T_s,d,r)$.
This probability determines the expected number of absorbed molecules, while the absorption locations determine the spatial structure used for detection.

Let \(\mathbf{x}_i\in\R^3\) denote the absorption position of the \(i\)th received molecule on the spherical Rx surface, so that \(\|\mathbf{x}_i\|=r\). The corresponding unit absorption direction is
\begin{equation}
    \mathbf{u}_i = \frac{\mathbf{x}_i}{r}, \qquad \mathbf{u}_i\in\Sph .
    \label{eq:unit_absorption_direction}
\end{equation}
The receiver also records the arrival-time bin \(b_i\in\{1,\ldots,B\}\). Hence, the observed data over the interval \([0,T_s]\) are
\begin{equation}
    \Dobs = \{(\mathbf{u}_i,b_i)\}_{i=1}^{\nrx},
    \label{eq:data}
\end{equation}
where \(\nrx\) is the number of absorbed molecules, and $b_i$ is the arrival-time bin associated with the $i^{th}$ absorbed molecule.

The detection of the system is a symbol-level binary test. Let \(x_n\in\{0,1\}\) denote the information symbol in the current observation interval. The hypotheses are
\begin{equation}
\begin{array}{ll}
    \Hnull: & x_n=0,\\[1mm]
    \Halt: & x_n=1 .
\end{array}
\label{eq:hypotheses}
\end{equation}
Here, \(\Hnull\) does not mean that the receiver observes no molecules, but means that Tx does not release an intended signal, so the observed molecules are explained by the noise component. This component may include environmental background molecules, unrelated molecular activity, and inter-symbol interference (ISI) from previous symbols.
We denote the normalized angular-time density of this expected noise by
\begin{equation}
    q_{\mathrm{N}}(\mathbf{u},b),
    \quad \text{where } \mathbf{u}\in\Sph \text{ and } b\in\{1,\ldots,B\}.
    \label{eq:noise_density}
\end{equation}
This definition over the surface and arrival-time bins satisfies
\begin{equation}
    \sum_{b=1}^{B}\int_{\Sph} q_{\mathrm{N}}(\mathbf{u},b)\,d\Omega = 1 .
    \label{eq:noise_normalization}
\end{equation}

The normalization in \eqref{eq:noise_normalization} means that
\(q_{\mathrm{N}}(\mathbf{u},b)\) describes the conditional angular-time
distribution of an observed noise/ISI molecule. To represent the number of such molecules, let \(\lambda_{\mathrm{N}}\) denote the expected number of molecules generated by this component in the current observation interval. Under \(\Hnull\), the per-molecule likelihood is \(p(\mathbf{u}_i,b_i\mid\Hnull)=q_{\mathrm{N}}(\mathbf{u}_i,b_i)\), and the count model is centered around \(\lambda_{\mathrm{N}}\).

Under \(\Halt\), the current symbol adds a structured component with expected count \(\lambda_{\mathrm{S}}\) and normalized angular-time density \(p_{\mathrm{S}}(\mathbf{u},b\mid d,\mathbf{a})\) (constructed in Section~\ref{sec_legendre}). The active-symbol likelihood is then the weighted superposition
\begin{equation}
    p(\mathbf{u}_i,b_i\mid\Halt,d,\mathbf{a})
    \!=\!
    \frac{\lambda_{\mathrm{N}}q_{\mathrm{N}}(\mathbf{u}_i,b_i)
    \!+\!\lambda_{\mathrm{S}}p_{\mathrm{S}}(\mathbf{u}_i,b_i\mid d,\mathbf{a})}
    {\lambda_{\mathrm{N}}+\lambda_{\mathrm{S}}},
\label{eq:p}
\end{equation}
with the count centered at \(\lambda_{\mathrm{N}}\!+\!\lambda_{\mathrm{S}}\).
The density \(q_{\mathrm{N}}\) can be learned from noise-only calibration data or built from an expected-ISI model: with environmental noise \(\lambda_{\mathrm{env}}q_{\mathrm{env}}\) and the \(m\)$^{th}$ previous symbol's expected contribution \(\lambda_m p_m(\mathbf{u},b)\),
\begin{equation}
    q_{\mathrm{N}}(\mathbf{u},b)
    =
    \frac{\lambda_{\mathrm{env}}q_{\mathrm{env}}(\mathbf{u},b)
    + \sum_{m=1}^{M-1}\lambda_m p_m(\mathbf{u},b)}
    {\lambda_{\mathrm{env}}+\sum_{m=1}^{M-1}\lambda_m},
\label{eq:expected_noise_distribution}
\end{equation}
where \(\lambda_{\mathrm{N}}=\lambda_{\mathrm{env}}+\sum_{m=1}^{M-1}\lambda_m\). Thus \(\Hnull\) models the noise-plus-ISI distribution, and \(\Halt\) adds the current ``1'' signal.

\section{Legendre polynomial angular-density approximation}
\label{sec_legendre}

After defining the geometry and observations, we now introduce the distributional model used for the structured current-signal component. For a Tx located in direction \(\mathbf{a}\in\Sph\), the angular coordinate of an absorption direction \(\mathbf{u}_i\) is
\begin{equation}
    z_i = \cos\theta_i = \mathbf{a}^{\mathsf{T}}\mathbf{u}_i,
    \qquad -1\leq z_i\leq 1 .
    \label{eq:z}
\end{equation}
Under the axial symmetry of the point-source Tx and spherical absorbing Rx geometry, the conditional angular distribution depends on \(\mathbf{u}_i\) only through \(z_i\). This reduces the surface-density modeling problem to a one-dimensional density approximation over \([-1,1]\).
Legendre polynomials \(P_\ell(z)\) form an orthogonal basis on \([-1,1]\), satisfying
\begin{equation}
    \int_{-1}^{1} P_\ell(z)P_m(z)\,dz
    = \frac{2}{2\ell+1}\indicator\{\ell=m\} .
    \label{eq:legendre_orthogonality}
\end{equation}
For a given distance \(d\) and arrival-time bin \(b\), let \(h(z\mid d,b)\) denote the density of the angular variable \(z\), conditioned on absorption in bin \(b\). We approximate this density by a finite Legendre expansion,
\begin{equation}
    h_L(z\mid d,b)
    =
    \sum_{\ell=0}^{L} c_\ell(d,b) P_\ell(z),
    \label{eq:legendre}
\end{equation}
where \(L\) is the approximation order and \(c_\ell(d,b)\) are the Legendre coefficients. If the exact density were available, the coefficients would be the orthogonal projections
\begin{equation}
    c_\ell(d,b)
    =
    \frac{2\ell+1}{2}
    \int_{-1}^{1} h(z\mid d,b)P_\ell(z)\,dz .
    \label{eq:legendre_coefficients}
\end{equation}
        In practice $h(z \mid d,b)$ is unknown, so the coefficients are estimated
from the particle-based samples of $z$ by a \yas{penalized least-squares fit to
the empirical angular histogram, with the ridge penalty
$\gamma \sum_{\ell=1}^{L} \ell^{2} c_{\ell}^{2}$, $\gamma = 10^{-6}$, to
suppress overfitting of high-order coefficients to sampling noise.}
Since the truncated expansion can take small negative values, we use the floored and renormalized density
\begin{equation}
    \bar h_L(z\mid d,b)
    =
    \frac{\max\{\epsilon,h_L(z\mid d,b)\}}
    {\int_{-1}^{1}\max\{\epsilon,h_L(s\mid d,b)\}\,ds},
    \label{eq:projected}
\end{equation}
where \(\epsilon>0\) is a numerical floor, so \(\int_{-1}^{1}\bar h_L(z\mid d,b)\,dz=1\).
The resulting surface density with respect to solid angle \(\Omega\) is
\begin{equation}
    f_{\Omega}(\mathbf{u}\mid d,\mathbf{a},b)
    =
    \frac{1}{2\pi}\bar h_L(\mathbf{a}^{\mathsf{T}}\mathbf{u}\mid d,b),
    \label{eq:surface_density}
\end{equation}
where \(z=\mathbf{a}^{\mathsf{T}}\mathbf{u}=\cos\theta\). Substituting \eqref{eq:surface_density} into the current-symbol model gives the normalized signal-component density
\begin{equation}
    p_{\mathrm{S}}(\mathbf{u}_i,b_i\mid d,\mathbf{a})
    =
    m_{b_i}(d)
    \frac{1}{2\pi}
    \bar h_L(\mathbf{a}^{\mathsf{T}}\mathbf{u}_i\mid d,b_i).
    \label{eq:p1_single}
\end{equation}
Since \(d\Omega=d\phi\,dz\), the factor \(1/(2\pi)\) makes \(f_\Omega\) normalized on \(\Sph\). With \(\sum_{b=1}^{B}m_b(d)=1\), \(p_{\mathrm{S}}\) is a normalized joint angular-time density on \(\Sph\times\{1,\ldots,B\}\), molecule counts are modeled separately through \(\lambda_{\mathrm{S}}=N_{\mathrm{Tx}}F_{\mathrm{hit}}(T_s,d,r)\).

To learn the angular hit density empirically, a particle-based MCvD simulator releases molecules from a point Tx at each distance $d$ on a grid $\mathcal{G}_d$ and records their absorption directions and arrival-time bins. Repeating \eqref{eq:z}--\eqref{eq:p1_single} for each tap $m = 0, \dots, K-1$ (the symbol interval of absorption) captures ISI and yields the reference library
\begin{equation}
\begin{split}
    \mathcal{L}_{\mathrm{Leg}}
    =\{&c_\ell^{(m)}(d,b),\,m_b^{(m)}(d),\,p^{(m)}_{\mathrm{cap}}(d):\\
    &d\in\mathcal{G}_d,\, m=0,\ldots,K-1,\\
    &b=1,\ldots,B,\,\ell=0,\ldots,L\}
\end{split}
\label{eq:library}
\end{equation}
stores, per tap, the angular coefficients \(c_\ell^{(m)}(d,b)\), the conditional time-bin masses \(m_b^{(m)}(d)\) with \(\sum_b m_b^{(m)}\!(d)\!=\!1\), and the per-tap absorption probability \(p^{(m)}_{\mathrm{cap}}\!(d)\!=\!F_{\mathrm{hit}}\!\big((m{+}1)T_s,d,r\big)\!-\!F_{\mathrm{hit}}\!\big(mT_s,d,r\big)\), so that \(p^{(0)}_{\mathrm{cap}}\!(d)\!=\!F_{\mathrm{hit}}\!(T_s,d,r)\). This library supplies the current-symbol density \(p_{\mathrm{S}}\) in the mixture likelihood \eqref{eq:p} and the per-tap intensities used by the detectors of Sec.~\ref{sec_methodology}, the quantities above are the current-symbol case \(m=0\).
In our experiments, \(L=10\), \(B=100\), \(r=\SI{5}{\micro\meter}\), \(D=\SI{79.4}{\micro\meter^2\per\second}\), \(d=\SI{12.5}{\micro\meter}\), and the simulation step is \(\SI{1e-4}{\second}\). No environmental noise is added, so the noise component is ISI only, and the memory length \(M\) (equivalently the number of taps \(K\)) is the smallest number of symbols for which \(F_{\mathrm{hit}}\) reaches \(75\%\) of its asymptotic value.

\section{Legendre Likelihood-Ratio Detection}
\label{sec_methodology}

\subsection{Legendre Detection Rule}
\label{subsec_single_lrt}

Given the observation set \(\Dobs\), the memoryless Legendre detector compares the expected-noise model under \(\Hnull\) with the noise-plus-current-signal model under \(\Halt\). Throughout, \(\lambda_m=N_{\mathrm{Tx}}p^{(m)}_{\mathrm{cap}}(d)\) denotes the expected count produced by a \emph{single} active emission \(m\) symbol intervals earlier. The expected current-signal count is then \(\lambda_{\mathrm{S}}=\lambda_0=N_{\mathrm{Tx}}p^{(0)}_{\mathrm{cap}}(d)\). Because the memoryless detector cannot observe the past bits, it replaces each unknown previous bit by its prior mean \(\bar{p}=P(x=1)\), giving the expected noise count \(\lambda_{\mathrm{N}}=\lambda_{\mathrm{env}}+\bar{p}\sum_{m=1}^{K-1}\lambda_m\). Correspondingly, the ISI part of \(q_{\mathrm{N}}\) weights the \(m\)th previous-symbol angular profile by the expected mass \(\bar{p}\lambda_m\). The environmental term \(q_{\mathrm{env}}(\mathbf{u},b)=\tfrac{1}{4\pi B}\) is uniform over the sphere and over the \(B\) time bins, with \(\lambda_{\mathrm{env}}=0\) in our experiments, \(\Hnull\) reduces to the prior-averaged expected-ISI angular distribution.
Using the Poisson counting model
\begin{equation}
    P_{\mathrm{N}}(\nrx;\lambda_{\mathrm{N}})
    =
    \frac{\lambda_{\mathrm{N}}^{\nrx}e^{-\lambda_{\mathrm{N}}}}{\nrx!},
    \label{eq:count_poisson_h0}
\end{equation}
\begin{equation}
    P_{\mathrm{N+S}}(\nrx;\lambda_{\mathrm{N}}+\lambda_{\mathrm{S}})
    =
    \frac{(\lambda_{\mathrm{N}}+\lambda_{\mathrm{S}})^{\nrx}
    e^{-(\lambda_{\mathrm{N}}+\lambda_{\mathrm{S}})}}{\nrx!},
    \label{eq:count_poisson_h1}
\end{equation}
the null and active-symbol log-likelihoods become
\begin{equation}
\begin{split}
    \!\!\!\ell_0(\Dobs)\!
    &\!=\!\!
    \log \!P_{\mathrm{N}}(\nrx;\lambda_{\mathrm{N}})
    \!+\!
    \sum_{i=1}^{\nrx}
    \log p(\mathbf{u}_i,\!b_i\!\mid\!\Hnull),
    \\
    \!\!\!\ell_1(\Dobs)\!
    &\!=\!\!
    \log \!P_{\mathrm{N\!+\!S}}(\nrx;\!\lambda_{\mathrm{N}}\!\!+\!\!\lambda_{\mathrm{S}})
    \!+\!\!
    \sum_{i=1}^{\nrx}
    \log p(\mathbf{u}_i,\!b_i\!\mid\!\Halt,\!d,\!\mathbf{a})
\end{split}
\label{eq:l0_l1}
\end{equation}

Each log-likelihood thus splits into a \emph{count} term and a \emph{shape} term. The likelihood-ratio statistic is \(\Lambda(\Dobs)=\ell_1(\Dobs)-\ell_0(\Dobs)\), and the detector decides \(\Halt\) if \(\Lambda(\Dobs)>\tau\), otherwise, it decides \(\Hnull\). The threshold \(\tau\) is selected to meet a desired BER or false-alarm probability. This count/shape decomposition is reused by the hybrid detector of Sec.~\ref{subsec_hybrid_legendre_viterbi}.

\subsection{Legendre-Viterbi Sequence Detection}
\label{subsec_legendre_viterbi}

We next extend the Legendre detector to sequence detection. The memoryless rule of Sec.~\ref{subsec_single_lrt} fixes the ISI background to its prior-averaged expectation, sequence detection removes this approximation by letting each candidate bit history specify exactly which past symbols were active. Let
\(\Dobs_n=\{(\mathbf{u}_{n,i},b_{n,i})\}_{i=1}^{N_n^{\mathrm{Rx}}}\)
denote the observations in the \(n\)th symbol interval, and let
\(\boldsymbol{\sigma}_{n-1}\) denote the Viterbi state containing the previous
\(K-1\) candidate bits. For a candidate transition with current bit
\(x_n\), the associated branch history is
\(
    (x_n,x_{n-1},\ldots,x_{n-K+1}),
\)
where the previous bits are specified by \(\boldsymbol{\sigma}_{n-1}\).
Each library tap defines the angular-time \emph{intensity} contributed by an active emission \(m\) symbol intervals earlier,
\begin{equation}
    \mu_m(\mathbf{u},b)
    =
    \underbrace{N_{\mathrm{Tx}}p^{(m)}_{\mathrm{cap}}(d)}_{=\,\lambda_m}\,
    m_b^{(m)}(d)\,
    f_\Omega^{(m)}(\mathbf{u}\mid d,\mathbf{a},b),
    \label{eq:mu_m_def}
\end{equation}
where \(f_\Omega^{(m)}(\mathbf{u}\mid d,\mathbf{a},b)=\tfrac{1}{2\pi}\bar{h}_L(\mathbf{a}^{\mathsf{T}}\mathbf{u}\mid d,b)\) is the normalized surface density of \eqref{eq:surface_density} evaluated with the tap-\(m\) coefficients \(c_\ell^{(m)}(d,b)\), and \(m_b^{(m)}(d)\) is the conditional time-bin mass \((\sum_b m_b^{(m)}=1)\). The product \(p_{\mathrm{S}}^{(m)}(\mathbf{u},b)=m_b^{(m)}(d)f_\Omega^{(m)}(\mathbf{u}\mid d,\mathbf{a},b)\) is the per-tap version of \eqref{eq:p1_single}, and \(\mu_m=\lambda_m p_{\mathrm{S}}^{(m)}\) is that density scaled by the per-tap expected count, so that \(\sum_b\int_{\Sph}\mu_m(\mathbf{u},b)\,d\Omega=\lambda_m=N_{\mathrm{Tx}}p^{(m)}_{\mathrm{cap}}(d)\). Adding all active taps in the candidate history, the total branch intensity is
\begin{equation}
    \mu_n(\mathbf{u},\!b\mid \!x_n,\boldsymbol{\sigma}_{n\!-\!1})
    \!=\!
    \lambda_{\mathrm{env}}q_{\mathrm{env}}(\mathbf{u},\!b)
    \!+\!
    \sum_{m=0}^{K\!-\!1}
    x_{n\!-\!m}\mu_m(\mathbf{u},\!b),
    \label{eq:branch_intensity}
\end{equation}
and the corresponding total expected number of absorbed molecules is
\begin{equation}
    \bar{\lambda}_n(x_n,\boldsymbol{\sigma}_{n-1})
    =
    \lambda_{\mathrm{env}}
    +
    \sum_{m=0}^{K-1}x_{n-m}\lambda_m .
    \label{eq:branch_total_intensity}
\end{equation}

The branch metric is obtained by modeling the absorbed marks
\(\{(\mathbf{u}_{n,i},b_{n,i})\}_{i=1}^{N_n^{\mathrm{Rx}}}\) as an inhomogeneous
Poisson point process on \(\Sph\times\{1,\ldots,B\}\) with intensity
\(\mu_n(\mathbf{u},b\mid x_n,\boldsymbol{\sigma}_{n-1})\). Since
\(\sum_b\int_{\Sph}\mu_n\,d\Omega=\bar{\lambda}_n\), the branch log-likelihood,
up to an additive constant independent of the candidate branch, is
\begin{equation}
\begin{split}
    \Gamma_n(x_n,\boldsymbol{\sigma}_{n-1})
    =
    &\log P(x_n)
    -
    \bar{\lambda}_n(x_n,\boldsymbol{\sigma}_{n-1})
    \\
    &+
    \sum_{i=1}^{N_n^{\mathrm{Rx}}}
    \log
    \mu_n(\mathbf{u}_{n,i},b_{n,i}\mid x_n,\boldsymbol{\sigma}_{n-1}) .
\end{split}
\label{eq:legendre_viterbi_metric}
\end{equation}
Thus, each Viterbi branch is scored using the angular-time intensity.
The Viterbi recursion is then
\begin{equation}
    M_n(\boldsymbol{\sigma}_n)
    =
    \max_{\boldsymbol{\sigma}_{n-1}\rightarrow\boldsymbol{\sigma}_n}
    \left[
    M_{n-1}(\boldsymbol{\sigma}_{n-1})
    +
    \Gamma_n(x_n,\boldsymbol{\sigma}_{n-1})
    \right],
    \label{eq:legendre_viterbi_recursion}
\end{equation}
followed by the standard traceback step. Equivalently,
\begin{equation}
    \hat{\mathbf{x}}
    =
    \arg\max_{\mathbf{x}\in\{0,1\}^{N}}
    \sum_{n=1}^{N}
    \Gamma_n(x_n,\boldsymbol{\sigma}_{n-1}).
    \label{eq:legendre_viterbi_decision}
\end{equation}

\subsection{Hybrid Legendre-Viterbi Detector}
\label{subsec_hybrid_legendre_viterbi}
The full detector \eqref{eq:legendre_viterbi_metric} re-evaluates the angular-time intensity for every candidate history, which couples the spatial likelihood to the trellis state. We also evaluate a lower-complexity \emph{hybrid} detector that decouples the two: it keeps the state-conditioned count likelihood of the count-based Viterbi recursion, but reuses a single precomputed Legendre shape term per symbol. For each symbol interval, we first compute
\begin{equation}
    \Delta_n^{\mathrm{sh}}
    =
    \ell_{1,\mathrm{sh}}(\Dobs_n)
    -
    \ell_{0,\mathrm{sh}}(\Dobs_n),
    \label{eq:hybrid_legendre_shape_llr}
\end{equation}
where \(\ell_{1,\mathrm{sh}}\) and \(\ell_{0,\mathrm{sh}}\) are the per-molecule angular-time shape log-likelihood terms of \eqref{eq:l0_l1}, with the Poisson count factors excluded. Hence \(\Delta_n^{\mathrm{sh}}\) is the shape part of the memoryless statistic of Sec.~\ref{subsec_single_lrt}, evaluated against the same prior-averaged expected-ISI background \(q_{\mathrm{N}}\), because that background does not depend on \(\boldsymbol{\sigma}_{n-1}\), \(\Delta_n^{\mathrm{sh}}\) is computed once per symbol and reused for every branch. The hybrid branch metric is
\begin{equation}
\begin{split}
    \Gamma_n^{\mathrm{hyb}}(x_n,\boldsymbol{\sigma}_{n\!-\!1})
    \!=
    &x_n\Delta_n^{\mathrm{sh}}
    \!+\!
    \log P(x_n)
    \\
    \!+\!&
    \log P\!\left(
        N_n^{\mathrm{Rx}};
        \bar{\lambda}_n(x_n,\boldsymbol{\sigma}_{n\!-\!1})
    \right) .
\end{split}
\label{eq:hybrid_legendre_viterbi_metric}
\end{equation}
The hybrid thus keeps the count likelihood fully state-conditioned, since ISI counts still enter \(\bar{\lambda}_n(x_n,\boldsymbol{\sigma}_{n-1})\) through the trellis, while collapsing the angular evidence into a single scalar that credits the angular signature of only the current active bit via \(x_n\Delta_n^{\mathrm{sh}}\). By contrast, the full detector \eqref{eq:legendre_viterbi_metric} injects every active tap's profile \(\mu_m\) into the intensity, so its angular term is genuinely state-dependent. 
Table~\ref{tab:complexity} summarizes the per-symbol costs. The hybrid detector retains the state-conditioned count recursion and one memoryless shape evaluation per symbol, whereas the full Legendre-Viterbi detector recomputes the angular-time intensity for each candidate branch.

\begin{table}[t]
\centering
\caption{Per-symbol computational complexity of detectors.}
\label{tab:complexity}
\begin{tabular}{lc}
\hline
Detector & Complexity \\
\hline
Fixed threshold & \(O(1)\) \\
Legendre & \(O(\nrx L)\) \\
Count Viterbi & \(O(2^{K-1})\) \\
Hybrid Legendre-Viterbi & \(O(2^{K-1}+\nrx L)\) \\
Full Legendre-Viterbi & \(O(2^{K}\nrx K)\) \\
\hline
\end{tabular}
\end{table}
%\vspace{-20em}

\section{Performance Evaluation}
\label{sec_performance}

\begin{figure}[!t]
    \centering
    \includegraphics[width=0.98\linewidth]{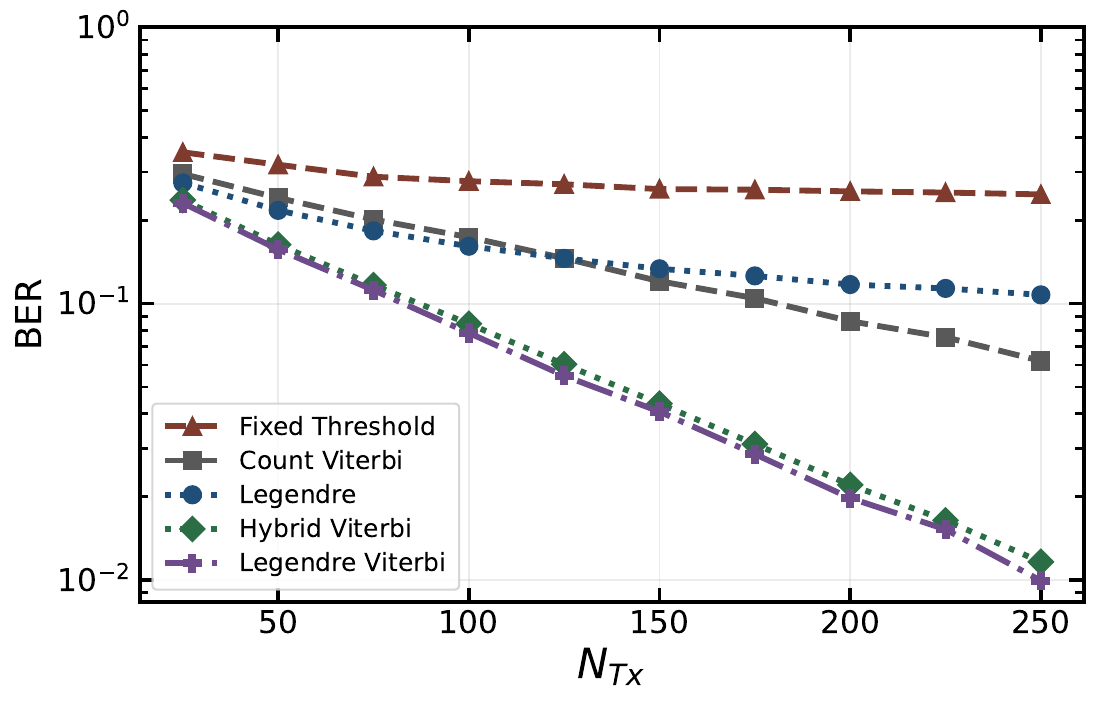}
\caption{BER versus \(N_{\mathrm{Tx}}\) for fixed symbol duration \(T_s=\SI{0.2}{\second}\), comparing detectors. Each point is estimated over \(100{,}000\) transmitted BCSK bits.}
    \label{fig:ntx}
\end{figure}

Fig.~\ref{fig:ntx} evaluates the BER as the molecule budget per active bit, $N_{\mathrm{Tx}}$, is varied while keeping the symbol duration fixed at \(T_s=\SI{0.2}{\second}\). As expected, increasing $N_{\mathrm{Tx}}$ reduces the BER for all detectors, since more released molecules improve the reliability of the received observations. The fixed-threshold detector performs the worst because it relies only on a scalar count comparison and cannot adapt to the statistical structure of the received signal. The count-based Viterbi detector benefits from sequence-level memory and provides a strong count-only baseline. The memoryless Legendre detector achieves comparable BER by exploiting the receiver-surface angular distribution, showing that the spatial distribution of absorption locations carries detection-relevant information beyond molecule counts. The Legendre-Viterbi detector further improves performance by combining this angular-time likelihood with Viterbi sequence memory, the gap between Count Viterbi and Legendre Viterbi quantifies the additional value of the receiver-surface hit pattern under sequence detection.

\begin{figure}[!t]
    \centering
    \includegraphics[width=0.98\linewidth]{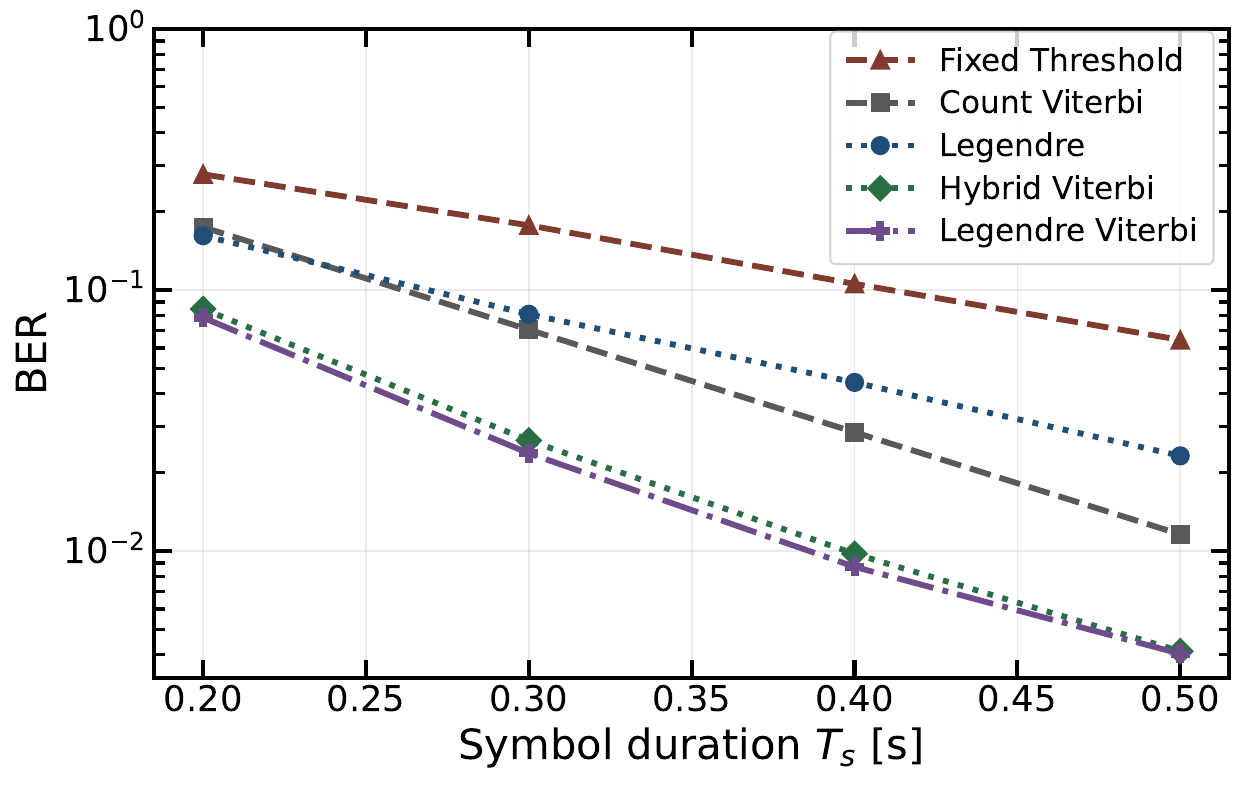}
    \caption{BER versus $T_s$ for a fixed molecule budget $N_{\mathrm{Tx}} = 100$. Shorter symbol durations correspond to a more difficult detection regime because fewer molecules are absorbed, and the count statistics become less reliable. The same BER evaluation procedure as in Fig.~\ref{fig:ntx} is used.}
    \label{fig:ts}
\end{figure}

Fig.~\ref{fig:ts} shows the BER as a function of the symbol duration $T_s$ for $N_{\mathrm{Tx}} = 100$. Longer symbol durations let more molecules arrive within each interval, improving the reliability of both count-based and spatial detectors. The advantage of the Legendre-based detectors is most visible in the small-$T_s$ regime, where the received molecule count is limited: the count-only information becomes less discriminative, whereas the Legendre detectors can still exploit the angular structure of the absorption pattern. The Legendre-Viterbi detector achieves the lowest BER because it uses both the spatial likelihood from the Legendre representation and the sequence-level memory of the Viterbi algorithm.

Together, these results demonstrate that receiver-surface angular information is useful in both memoryless and sequence-aware detection. The memoryless Legendre detector matches count-based Viterbi without explicit memory across symbols, and combining the same Viterbi principle with the Legendre angular-time likelihood yields the best performance, especially in low-count regimes such as small $T_s$ or small $N_{\mathrm{Tx}}$.

\begin{figure}[!t]
    \centering
    \includegraphics[width=0.98\linewidth]{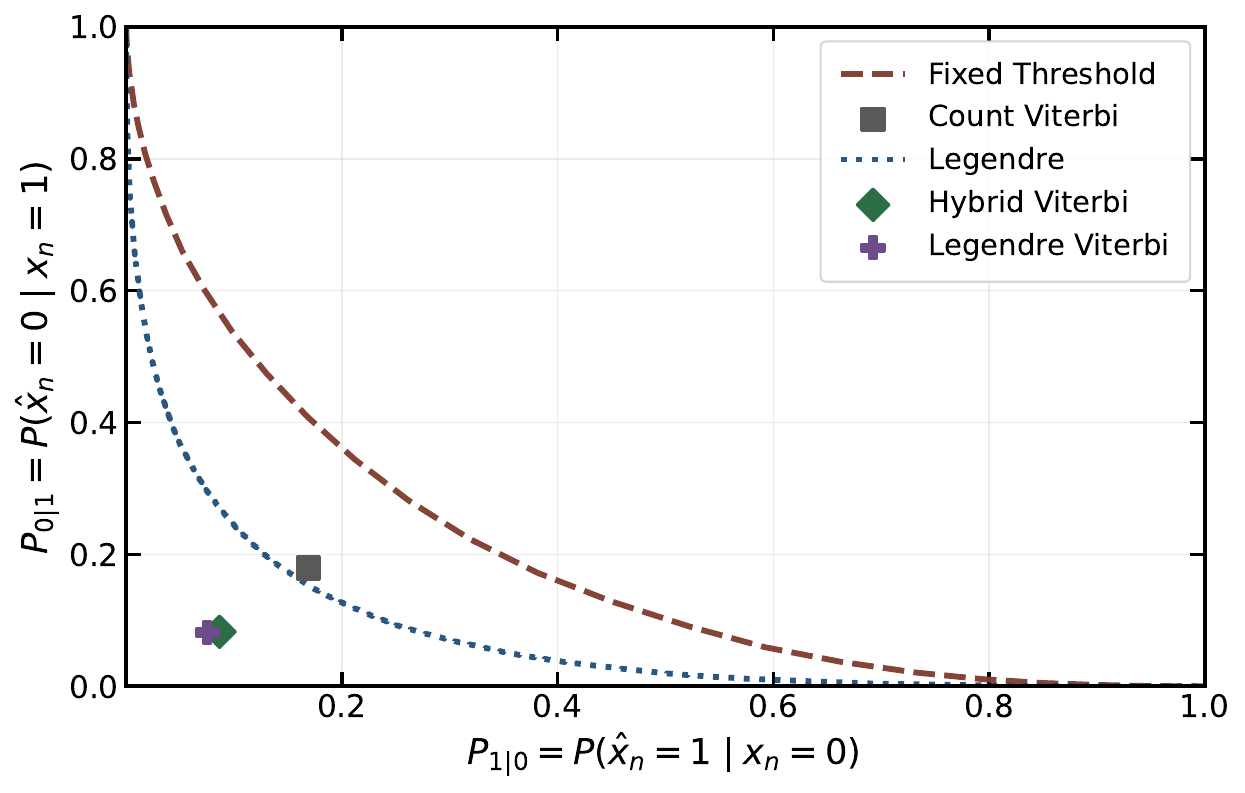}
    \caption{\(P_{0|1}\) versus \(P_{1|0}\) for \(N_{\mathrm{Tx}}=100\) and \(T_s=\SI{0.2}{\second}\). Lower-left operating points indicate better detection performance, corresponding to simultaneously lower false-alarm and missed-detection probabilities.}
    \label{fig:pfa_pmiss}
\end{figure}

Fig.~\ref{fig:pfa_pmiss} compares the detectors in terms of \(P_{0|1}\) and \(P_{1|0}\): the horizontal axis is \(P_{1|0}\), the probability of deciding the current symbol is \yas{bit-1} when %\(\Hnull\) is true
\yas{it is bit-0}, and the vertical axis is \(P_{0|1}\), the probability of %missing an active symbol when \(\Halt\) is true.
\yas{deciding bit-0, when it is bit-1.}
Detectors closer to the lower-left corner provide a better %false-alarm/missed-detection balance
\yas{detection performance}. The fixed-threshold detector exhibits the weakest tradeoff because it relies only on the molecule count, the Legendre detector improves it by using the angular-time distribution of the absorbed molecules.
\section{Conclusion}
\label{sec_conclusion}
This \yas{letter} introduced a Legendre-polynomial-based detection method for MCvD with a perfectly absorbing spherical receiver. By modeling the angular distribution of molecule absorption locations, the proposed detector exploits receiver-surface information that is discarded by count-only methods. This matters because count-only detectors can fail when a weak structured signal and a stronger background process yield similar values of \(\nrx\), whereas the Legendre detector compares the observed angular pattern under two physically meaningful models. 
A true \yas{bit-1} symbol shifts the angular-time pattern toward the Legendre profile around direction \(\mathbf{a}\). This profile varies with distance and time and is encoded compactly by the Legendre coefficients. As a result, the likelihood-ratio test respects the curvature and axial symmetry of the absorption process.

Across the evaluated regimes, the memoryless Legendre detector achieves performance comparable to the count-based Viterbi detector without Viterbi's exponential sequence complexity. The hybrid Legendre-Viterbi detector combines state-conditioned count likelihoods with a memoryless Legendre shape term, while the full Legendre-Viterbi detector achieves the best BER by using state-dependent angular-time likelihoods with sequence memory. The gains are especially visible in \yas{short $T_s$} regimes, where count statistics become less reliable. Future work can extend this framework to multi-transmitter scenarios, joint detection and localization, and receivers with partial or noisy surface-position measurements.
\bibliographystyle{IEEEtran}
\bibliography{refs_final}

\end{document}